\documentstyle[12pt,aps,prl,psfig]{revtex}
\begin{document}
\draft
\title{Perturbation Theory for Quantum Computation \\
with Large Number of Qubits}
\author{G.P. Berman$^1$, G.D. Doolen$^1$, D.I. Kamenev$^1$,
and V.I. Tsifrinovich$^2$}
\address{$^1$ Theoretical Division and CNLS, Los Alamos National Laboratory,
Los Alamos, NM 87545}
\address{$^2$ IDS Department, Polytechnic University, Six Metrotech
Center, Brooklyn, New York 11201}
\maketitle
\begin{abstract}
We describe a new and consistent perturbation theory for solid-state
quantum computation with many qubits. The errors
in the implementation of simple quantum
logic operations caused by non-resonant transitions are estimated.
We verify our perturbation approach using
exact numerical solution for relatively small ($L=10$)
number of qubits. A preferred range of parameters is found in which
the errors in processing quantum information are small.
Our results are needed for experimental testing of scalable solid-state
quantum computers.
\end{abstract}

Several proposals for scalable solid-state quantum computers have been
recently published \cite{10,kane,11,12,13,14,14.1,divincenzo,mrfm,1.1,1.2}.
For the most effective quantum information
processing all of these proposals require operations with large number of qubits.
In Ref.~\cite{1} quantum logic operations between remote qubits
were simulated in a chain of $L=1000$ nuclear spins.
Such a quantum problem can be solved only using approximate methods
since the dimensionality of the Hilbert space increases as $2^L$.
The approximate procedure used in \cite{1}
was based on the selection of states generated as a result
of resonant or near-resonant transitions, while the other transitions were
neglected in a controlled way.
In this Letter we develop a consistent approach to this problem
based on perturbation theory. We use our procedure to analytically estimate
the probability of generation of unwanted
states caused by non-resonant transitions and
verify that these agree with the exact numerical solution
for the same problem with relatively small number of qubits ($L=10$),
for which the dimensionality of the Hilbert space is not very large
($N=2^{10}=1024$).

{\it Dynamics of a spin chain}
We consider a chain of identical nuclear spins placed in an external
high magnetic
field, $B(x)$, which has uniform gradient in the direction of the chain, $x$.
The nuclear magnetic resonance (NMR) frequency for the $k$th spin
is $\omega_k=\gamma B_k$, where
$\gamma$ is the nuclear gyromagnetic ratio and $B_k$ is the
$z$-component of the
magnetic field at the location of the $k$th spin.
The gradient of the magnetic field
provides a change in $\omega_k$ by the value $\delta\omega$
between the neighboring spins. (For physical parameters see
Ref.~\cite{1}.)

The Hamiltonian of the spin chain in an external radio-frequency (rf)
field is,
$$
H^{(n)}=-\sum_{k=0}^{L-1}\omega_kI_k^z-2J\sum_{k=0}^{L-1}I_k^zI_{k+1}^z-
$$
\begin{equation}
\label{H}
\Theta_n(t)(\Omega_n/2)
\sum_{k=0}^{L-1}[I_k^-\exp\left(-i(\nu_n t+\varphi_n\right))+
I_k^+\exp(i(\nu_n t+\varphi_n))]=
H_0+V^{(n)}(t),
\end{equation}
where $J$ is the Ising interaction constant between neighboring spins
and $I_k^z$ is the operator
of the $z$-component of spin $1/2$;
$\Omega_n$, $\nu_n$ and $\varphi_n$ are the Rabi frequency,
the frequency and the phase of the $n$th pulse,
$I_k^\pm=I_k^x\pm I_k^y$,
and $\Theta_n(t)$ equals 1 only during the $n$th pulse.

In this Letter we estimate the errors generated during the creation
of the entangled state for remote qubits
from the ground state of the chain by applying a single $\pi/2$ pulse
and a sequence of $\pi$ pulses in the system described by the
Hamiltonian (\ref{H}). First, we describe schematically the quantum protocol.
The first $\pi/2$ pulse creates a superposition of two states with equal
probabilities,
$|00\dots00\rangle\rightarrow (1/\sqrt 2)(|00\dots00\rangle+
i|10\dots00\rangle)$. The subsequent pulses, which we describe by the the
unitary transformation, $U$, transform this state
to an entangled state for remote qubits,
\begin{equation}
\label{U}
U\frac 1{\sqrt 2}(|00\dots00\rangle+
i|10\dots00\rangle)=\frac 1{\sqrt 2}(e^{i\phi_1}(|00\dots00\rangle+
e^{i\phi_2}|10\dots 01\rangle),
\end{equation}
where $\phi_1$ and $\phi_2$ are known phases (see Ref. \cite{1}).
The operator $U$ realizes
a particular case of the well-known CONTROL-NOT gate.
It has the following properties:
\begin{equation}
\label{U1}
U\frac 1{\sqrt 2}|0_{L-1}0_{L-2}\dots 0_{1}0_0\rangle=
\frac 1{\sqrt 2}e^{i\phi_1}|0_{L-1}0_{L-2}\dots 0_10_0\rangle,
\end{equation}
\begin{equation}
\label{U2}
U\frac 1{\sqrt 2}|1_{L-1}0_{L-2}\dots 0_{1}0_0\rangle=
\frac 1{\sqrt 2}e^{i\left(\phi_2-\frac\pi 2\right)}
|1_{L-1}0_{L-2}\dots 0_11_0\rangle.
\end{equation}
To accomplish the operation (\ref{U2}) in the
system described by the
Hamiltonian (\ref{H}), we choose a sequence of $\pi$ pulses with resonant
frequencies which will be described elsewhere.
If we apply the same protocol to the ground state (operation (\ref{U1})),
then with some probability the system will remain in this state because
these pulses have the detunings from resonant transitions,
$\Delta_n\ne 0$. The
near-resonant transitions have the probabilities \cite{1},
\begin{equation}
\label{epsilon}
\varepsilon_n=(\Omega_n/\lambda_n)^2\sin^2(\lambda_n \tau_n/2),
\end{equation}
where $\lambda_n=\sqrt{\Delta_n^2+\Omega_n^2}$ is the effective
field in frequency units and
$\tau_n$ is the duration of the $n$th pulse.
The values of the detunings are the same for all
pulses, $\Delta_n=\Delta=2J$,
except for the fourth
pulse, where $\Delta_4=4J$.
In our calculations we assumed
$\Omega_n=\Omega$ for $n\ne 4$ and $\Omega_4=2\Omega$, so that
for all pulses
the values
of $\varepsilon_n$ are the same, $\varepsilon_n=\varepsilon$.

We write the wave function, $\Psi(t)$, in the time-interval of the $n$th pulse,
in the laboratory system of coordinates
in the form,
\begin{equation}
\label{Psi1}
\Psi(t)=
\exp\left[i(\nu_nt+\varphi_n)\sum_{k=0}^{L-1}I_k^z\right]\Psi_{rot}(t)=
\sum_pA_p(t)|p\rangle\exp(-i\chi_p^{(n)}t+\xi_p^{(n)}),
\end{equation}
where $\Psi_{rot}(t)$ is the wave function in a frame rotating with
the frequency $\nu_n$.
$\chi_p^{(n)}=-(\nu_n/2)\sum_{k=0}^{L-1}\sigma_k^p$,
$\xi_p^{(n)}=\varphi_n\sum_{k=0}^{L-1}\sigma_k^p$,
$\sigma_k^p=-1$ if the $k$th spin of the state $|p\rangle$
is in the position
$1$ and $\sigma_k^p=1$ if the $k$th spin is in the position $0$,
$|p\rangle$ is the eigenfunction of the Hamiltonian $H_0$.
The dynamics
during the $n$th pulse with the frequency $\nu_n$ is described
by the following Schr\"odinger equation for the coefficients $A_p(t)$,
\begin{equation}
\label{Sch2}
i\dot A_p(t)=(E_p-\chi_p^{(n)})A_p(t)-\frac\Omega 2\sum_{p'}A_{p'}(t),
\end{equation}
where the sum is taken over the states $|p'\rangle$
connected by a single-spin transition with the state $|p\rangle$.
Eq. (\ref{Sch2}) can be written in the form
$i\dot A_p(t)={\cal H}_{pp'}^{(n)}A_{p'}(t)$, where
${\cal H}_{pp'}^{(n)}=H_{pp'}^{(n)}-\chi_p^{(n)}\delta_{p,p'}$, where
$\delta_{p,p'}$ is the Kronecker $\delta$-symbol.

Under the condition $\Delta\ll \delta\omega$
the energy separation between the $p$th and the $m$th
diagonal elements of the matrix $H_{pp'}^{(n)}$
connected by the resonant or near-resonant transition is
$|{\cal E}_{p}^{(n)}-{\cal E}_{m}^{(n)}|=\Delta_{n}^{pm}$,
where ${\cal E}_p^{(n)}=E_p-\chi_p^{(n)}$.
This is much less than the energy separation
between these diagonal elements and diagonal elements of the other states
connected with the states $|p\rangle$ and $|m\rangle$
by the non-resonant transitions.
In this case
one can neglect the interaction of the $p$th state with
all states except the state $|m\rangle$.
In this approximation the Hamiltonian matrix
${\cal H}_{pp'}^{(n)}$ breaks up into $N/2$
approximately independent matrices $2\times 2$, where $N=2^L$.

{\it Errors in the creation of an entangled state for remote qubits.}
The matrix approach allows us to estimate errors in the logic gate (\ref{U})
caused by near-resonant and
non-resonant transitions. Suppose that initially
the eigenstate $|p\rangle$ of the Hamiltonian $H_0$
is populated. We want to calculate
the probability of non-transition to the state $|r\rangle$ with
$|E_p-E_{r}|\sim \delta\omega$, where the states $E_p$ and $E_{r}$ are
connected by a flip of $k'$th spin, whose NMR frequency differs
by the value $\sim \delta\omega$ from the frequency
of the resonant transition. Since the matrix
elements are small, we can write
\begin{equation}
\label{psi10}
\psi^{(n)}_{q}=\psi^{0\,(n)}_{q}+
{\sum_{p}}'{v_{qp}^{(n)}\over {\cal E}^{(n)}_q-{\cal E}^{(n)}_{p}}
\psi^{0\,(n)}_{p},
\end{equation}
where prime in the sum means that the term with $p=q$ is omitted;
$v_{qp}^{(n)}$ is the matrix element for transition between
the states $\psi^{0\,(n)}_{q}$ and $\psi^{0\,(n)}_{p}$.

Because the matrix ${\cal H}_{qp}^{(n)}$ is divided into
$2^{L-1}$ relatively independent $2\times 2$ blocks,
the wave function, $\psi^{0\,(n)}_{p}$,
in Eq. (\ref{psi10}) is an eigenfunction of a single
block $2\times 2$ with all other elements being equal to zero.
In the case for which the two states $|p'\rangle$ and
$|p\rangle$ are connected by a near-resonant
transition, the function,
$\psi^{0\,(n)}_{p}$, in Eq. (\ref{psi10})
has the form, $\psi^{0\,(n)}_{p}\approx
[1-(\Omega^2/8\Delta^2)]|p\rangle+
(\Omega/2\Delta)|p'\rangle$. On the other hand,
if the states $|p\rangle$ and $|p'\rangle$
are connected by an exact resonant transition, we have
$\psi^{0\,(n)}_{p}=(1/\sqrt 2)(|p\rangle+|p'\rangle)$ and
$\psi^{0\,(n)}_{p'}=(1/\sqrt 2)(|p\rangle-|p'\rangle)$.
In both cases
the probability of non-resonant transition from the state
$|q\rangle$ to the state $|p\rangle$ connected with the state $|q\rangle$
by flip of the $k'$th spin (up to the value
$\mu=(\Omega/\omega)^2$) is
\begin{equation}
\label{psi9}
P_{pq}^{(n)}=\left|\langle p|\psi_q\rangle\right|^2=
\left({V_{pq}^{(n)}\over {\cal E}_q^{(n)}-{\cal E}_{p}^{(n)}}\right)^2\approx
\left({\Omega_n\over 2|k_n-k'|\delta\omega}\right)^2,
\end{equation}
where $|k_n-k'|$ is the distance from the $k'$th spin (whose NMR frequency,
$\omega_{k'}$, is non-resonant) to the $k_n$th spin with resonant
(or near-resonant) NMR frequency.

The probability
$\mu_{L-1}$ (here the subscript of $\mu$ stands for the number of the
resonant spin)
of generation of unwanted states by the first $\pi/2$ pulse
in the result of non-resonant transitions is
\begin{equation}
\label{Pnr1}
\mu_{L-1}=\left(\Omega\over 2\delta\omega\right)^2\sum_{k'=0}^{L-2}\frac 1{|L-1-k'|^2}.
\end{equation}
After the first $\pi/2$ pulse, the probability of the correct procedure
in implementation of the logic gate is $P_1=1-\mu_{L-1}$.
The probability of correct implementation of the operation (\ref{U})
by applying $2L-2$ pulses is
$$
P_{2L-2}=\frac 12(1-\mu_{L-1})(1-\mu_{L-2}-\varepsilon)
(1-4\mu_{L-2}-\varepsilon)(1-\mu_0-\varepsilon)
\prod_{i=1}^{L-3}(1-\mu_i-\varepsilon)^2+
$$
\begin{equation}
\label{Pnr}
\frac 12(1-\mu_{L-2})(1-4\mu_{L-2})\prod_{i=0}^{L-3}(1-\mu_i)^2,
\end{equation}
where the first term is provided by the
operation (\ref{U1}), and the
last term is due to (\ref{U2}).

\begin{figure}
\centerline{\mbox{\psfig{file=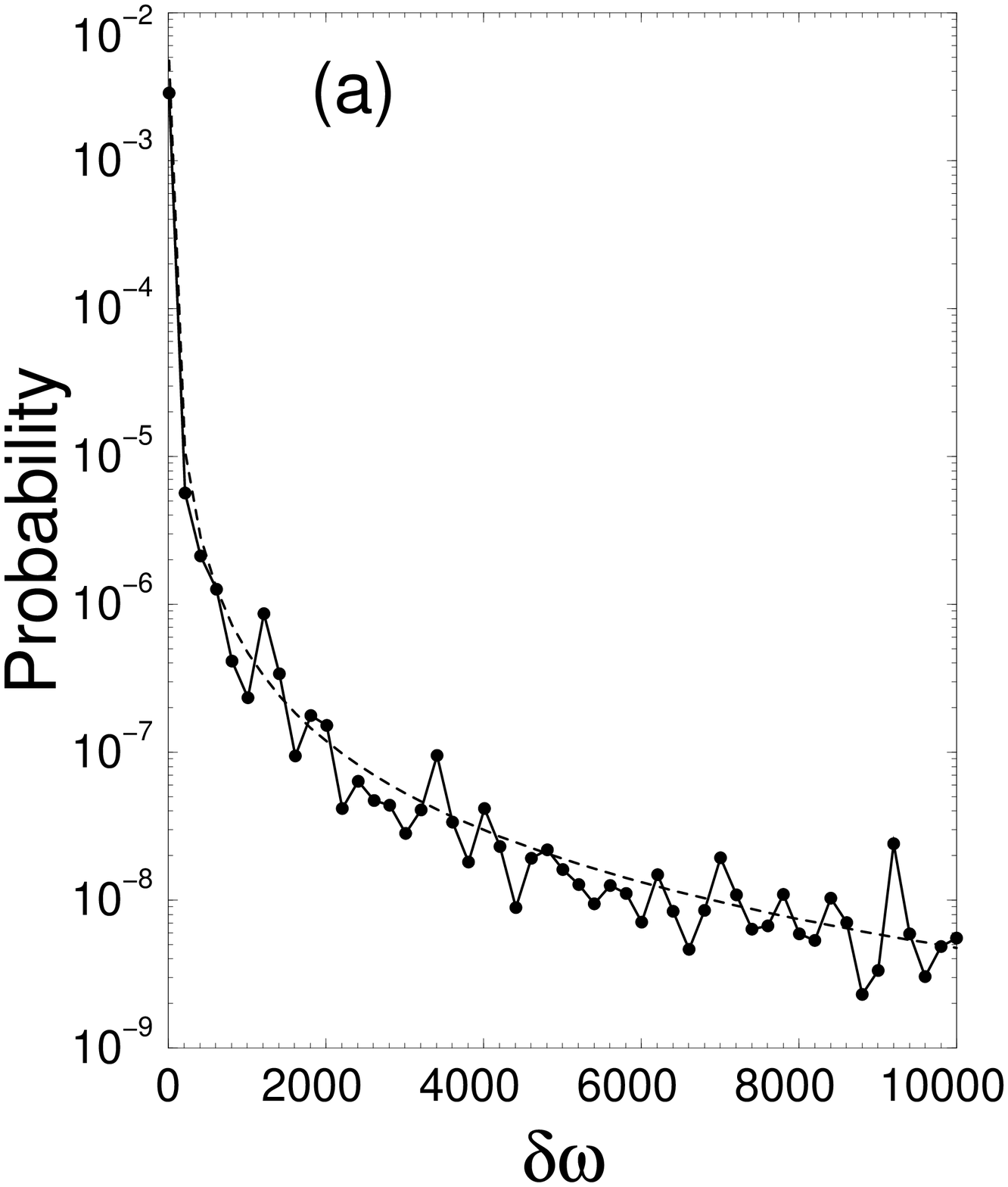,width=7cm,height=7cm}
       \psfig{file=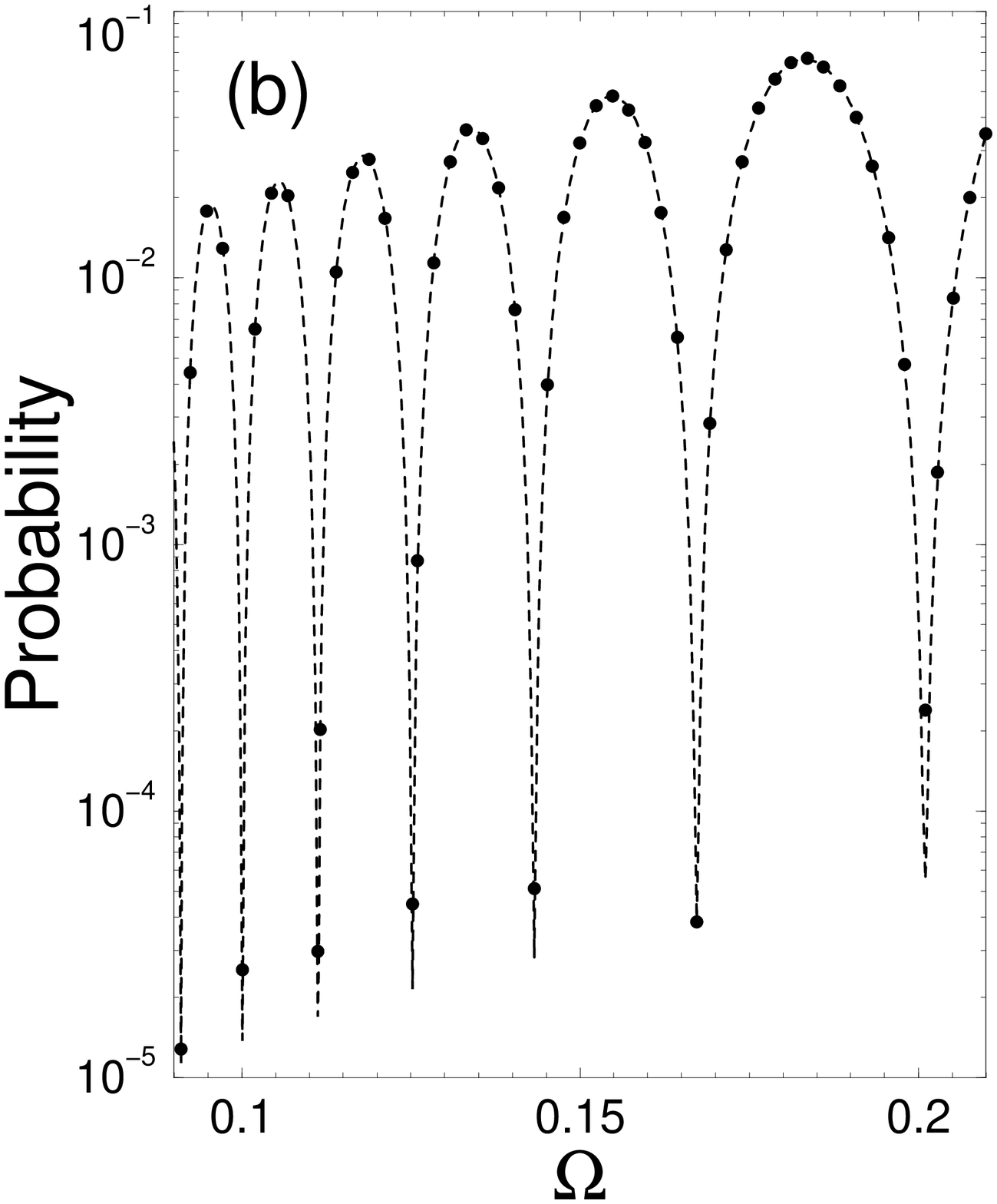,width=7cm,height=7cm}}}
\caption{(a) The probability, ${\cal P}$,
of generation of unwanted states in implementation of the logic gate
(\ref{U}) at $\varepsilon=0$. Filled
circles connected by the solid line are the numerical results,
dashed line is the analytic estimate;
$\Omega=2J/\sqrt{(4k^2-1)}$, $k=8$.
(b) ${\cal P}$ as a function of $\Omega$. Filled
circles are the numerical results.
The dashed line is the analytic estimate,
$\delta\omega=100$; $J=1$, $L=10$, $\varphi_n=0$ for all $n$.}
\label{fig:1}
\end{figure}

{\it Numerical results.}
In Fig. 1 (a) we compare the
total probability, ${\cal P}\equiv 1-P_{2L-2}$,
of generation of unwanted states with the result of exact numerical
solution of Eq. (\ref{Sch2}) for the case in which
the probability of near-resonant
transitions is negligibly small, i.e. when
$\varepsilon=0$. From Fig.~1~(a) one can see that
in this situation ${\cal P}$ decreases as
$\delta\omega$ increases.

When $\varepsilon$ is large, $\varepsilon\gg\mu$,
and $\Delta\ll\delta\omega$, the probability of generation of unwanted
states is mostly defined
by the value of $\varepsilon$
and is almost independent of $\delta\omega$.
In Fig. 1 (b) we plot the probability
${\cal P}$ as a function of $\Omega$. The value
of $\delta\omega$ fixes the values of the minima, ${\cal P}_{min}$,
in Fig. 1 (b): for larger
$\delta\omega$ the minima in the plot in Fig. 1 (b) become deeper.
The values of different minima in Fig. 1 (b) indicate the contribution
of non-resonant processes to the
probability ${\cal P}$. Since the value of $\Omega$ in Fig. 1 (b)
does not change significantly, the contribution of non-resonant
processes to the probability of errors is approximately the same for all
$\Omega$ and is equal to ${\cal P}_{min}$.
One can see that this contribution is negligibly small in comparison
with the contribution of near-resonant processes (defined by $\varepsilon$)
for all $\Omega$, except for the small regions of $\Omega$, where
$\varepsilon$, defined by Eq. (\ref{epsilon}), is minimal.

\begin{figure}
\centerline{\mbox{\psfig{file=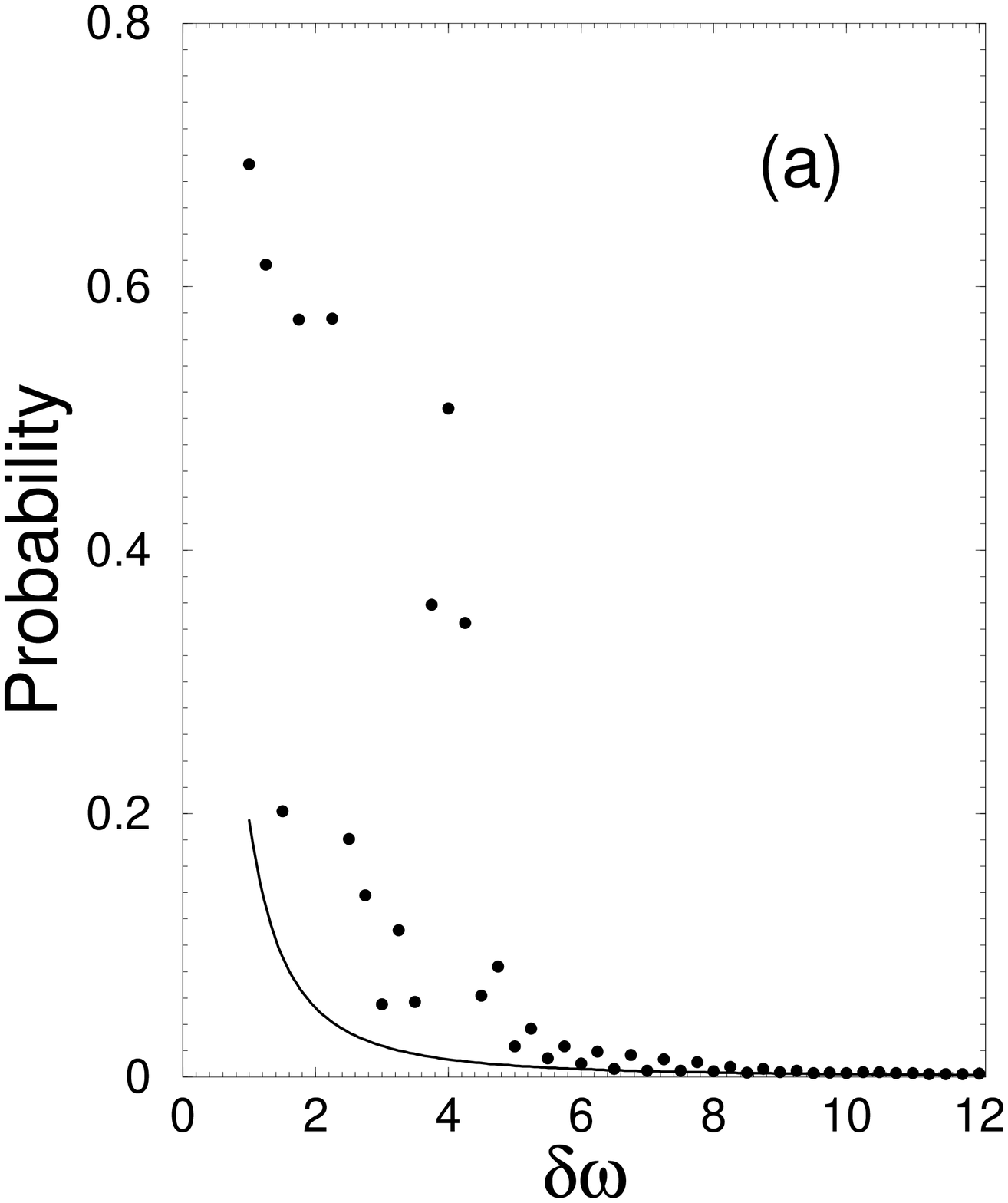,width=7cm,height=7cm}
       \psfig{file=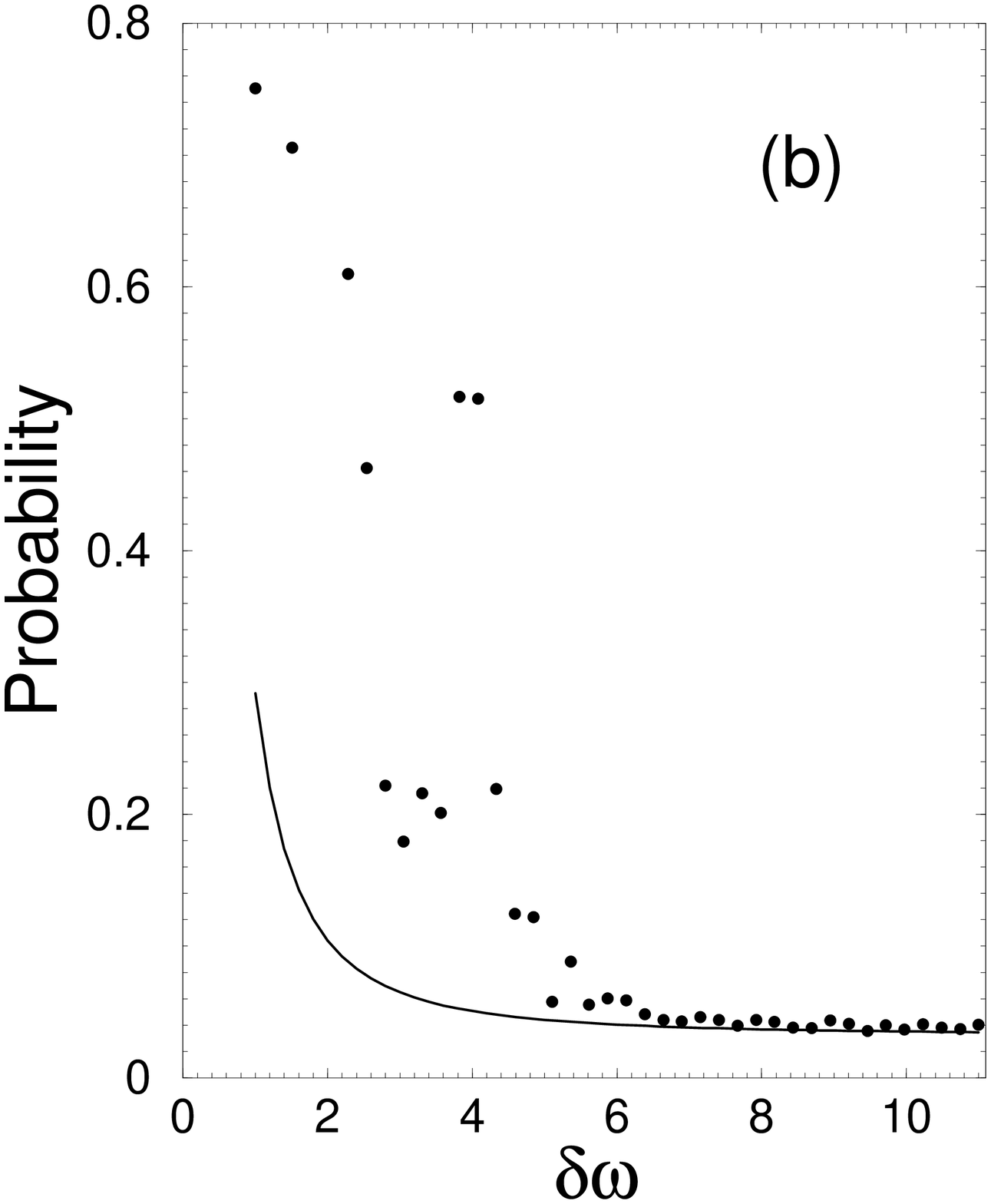,width=7cm,height=7cm}}}
\caption{The total probability of unwanted states
in implementation of the logic gate
(\ref{U}) as a function of $\delta\omega$ when the value of
$\delta\omega$ is comparable with the value of the detuning, $\Delta$.
(a) $\Omega=2J/\sqrt{4k^2-1}$, k=8 ($\varepsilon=0$), (b) $\Omega=0.15$
($\varepsilon=0.0039$).
Filled circles are the numerical results,
solid line is the analytical estimate for ${\cal P}$;
$J=1$, $L=10$, $\varphi_n=0$ for all $n$.}
\label{fig:4}
\end{figure}

When the perturbation parameters of the problem, $\varepsilon$ and $\mu$,
are small, using Eq. (\ref{Pnr}) one can easily estimate the probability
of generation of unwanted states in the implementation of the logic gate
(\ref{U}) when the
number of qubits in the spin chain is large (for example, for $L\sim 1000$).
In this case the Eq. (\ref{Pnr}) is important for estimation of errors,
since the exact solution of the problem requires
diagonalization of enormous matrices of size $2^L\times 2^L$.

We should note that one more condition (except for $\varepsilon,\,\mu\ll 1$)
must be satisfied  for Eq. (\ref{Pnr}) to be valid. The value
of detuning, $\Delta$, should be small in comparison with the
difference between NMR frequencies of the spins, $\Delta\ll\delta\omega$.
In Figs. 2 (a) and (b) we plot
the probability, ${\cal P}$, as a function of
$\delta\omega$ for small and large values of $\varepsilon$.
One can see that our results are valid
only when $\Delta\ll\delta\omega$, in spite of the fact that the
parameter $\Delta/\delta\omega$ does not appear explicitly
in Eq. (\ref{Pnr}). From Fig. 2 (b) one can see that the probability
of unwanted states, ${\cal P}$ for $\varepsilon\gg\mu$ and for large
$\delta\omega$ becomes
relatively independent of $\delta\omega$. In this case the value
of ${\cal P}$
is defined by $\varepsilon$ which, due to Eq. (\ref{epsilon}), does not
depend on $\delta\omega$.

This work was supported by the Department of Energy (DOE) under
contract W-7405-ENG-36, by the National Security Agency (NSA), and by the
Advanced Research and Development Activity (ARDA).

{}
\end{document}